\documentclass[11pt,a4paper]{article}

\usepackage[utf8]{inputenc}
\usepackage[T1]{fontenc}
\usepackage{amsmath,amssymb}
\usepackage{graphicx}
\usepackage{booktabs}
\usepackage{hyperref}
\usepackage[margin=2.0cm]{geometry}
\usepackage{xcolor}

\usepackage{algorithmic}
\usepackage{algorithm}
\usepackage{multirow}
\usepackage{enumitem}
\usepackage{float}
\usepackage{caption}
\usepackage{subcaption}
\usepackage{fancyhdr}
\usepackage{titlesec}

\titlespacing*{\section}{0pt}{12pt plus 3pt minus 2pt}{6pt plus 2pt minus 1pt}
\titlespacing*{\subsection}{0pt}{10pt plus 2pt minus 2pt}{4pt plus 1pt minus 1pt}
\titlespacing*{\paragraph}{0pt}{8pt plus 2pt minus 1pt}{0.5em}

\title{No\=esis: Bidirectional Graph-RAG with Adaptive Parallelism\\and Cross-Knowledge-Base Semantic Discovery\thanks{Patent pending. Application No.~102026000023146; Application No.~IT202500035167.}}

\author{Nicola Cogotti\\Alpha Cogs\\
\texttt{nicola.cogotti@alphacogs.co.uk}}

\date{August 2026}

\begin{document}
\maketitle

\begin{abstract}
Retrieval-Augmented Generation over knowledge graphs (Graph-RAG) has emerged as a powerful paradigm for grounding large language models in domain-specific corpora. However, existing systems face persistent limitations: (1)~static chunking fragments long documents, losing cross-section semantic connections; (2)~ingestion pipelines do not scale adaptively, causing out-of-memory failures or underutilized hardware; and (3)~multi-domain deployments require either a monolithic knowledge base (KB) that dilutes retrieval precision or manual user routing.

We present \textbf{No\=esis}, a decoupled Graph-RAG architecture that addresses these limitations through four novel algorithms: (a)~\emph{Bidirectional Graph Traversal} with a Graph-Feedback Context Resolver that simulates human sequential reading with degrading memory, producing significantly denser graphs versus single-pass baselines; (b)~an \emph{AIMD Concurrency Controller} adapted from TCP congestion control to RAG pipeline orchestration, achieving $23\times$ speedup with zero OOM events observed; (c)~\emph{Mo\=esis}, a domain-aware selective quantization pipeline for Mixture-of-Experts models that achieves $6.3\times$ prompt processing speedup on 12\,GB consumer GPUs and maintains stability even under extreme memory constraints; and (d)~\emph{Mesh}, a cross-KB semantic routing system with runtime structural discovery and adaptive Natural Break thresholds that enables small on-premises models to perform multi-hop cross-domain reasoning---discovering connections not explicitly present in any single document---as demonstrated in our evaluation.

We report measured results on real-world workloads: 1\,min\,6\,s for 13.4\,MB corpus ingestion (vs.\ 25\,min sequential), prompt processing speedup of $6.3\times$ on consumer GPUs, and cross-KB routing latency under 2\,ms. On the HotpotQA multi-hop QA benchmark (1{,}000 questions), No\=esis achieves 59.5 EM / 74.7 F1---surpassing GraphRAG by +27.8 EM while using a 35B on-premises model for graph construction rather than GPT-4o. Source text verification on a 193-page document confirms 90\% precision on extracted long-range causal edges, demonstrating that the bidirectional traversal captures cross-section relationships inaccessible to chunk-independent extraction.
\end{abstract}

\section{Introduction}
\label{sec:intro}

Knowledge graphs provide structured representations of domain expertise that can ground large language model (LLM) responses in verifiable facts. Graph-based Retrieval-Augmented Generation (Graph-RAG) combines the semantic richness of knowledge graphs with the generative capabilities of LLMs, enabling multi-hop reasoning over complex corpora~\cite{edge2024graphrag,guo2024lightrag}.

Despite recent progress, production deployment of Graph-RAG systems faces three critical bottlenecks:

\paragraph{Semantic Fragmentation.} Existing Graph-RAG systems---including Microsoft GraphRAG~\cite{edge2024graphrag}, LightRAG~\cite{guo2024lightrag}, LazyGraphRAG~\cite{lazygraphrag2024}, and HippoRAG~\cite{gutierrez2024hipporag}---extract entities and relations from documents using \emph{static chunking}: the document is split into fixed-size blocks, each processed independently. Limited or no shared context propagates between chunks during extraction. Consequently, concepts spanning chunk boundaries are lost or duplicated, and the resulting graph lacks the density required for reliable multi-hop traversal.

\paragraph{Rigid Parallelism.} Ingestion pipelines either process documents sequentially (Microsoft GraphRAG users report indexing times of hours to days for medium corpora~\cite{edge2024graphrag}) or use fixed parallelism that risks out-of-memory (OOM) crashes on heterogeneous hardware. To the best of our knowledge, no existing RAG ingestion pipeline adapts its document-level concurrency at runtime with persistent cross-worker state and crash recovery.

\paragraph{Cross-Domain Isolation.} Organizations maintaining multiple specialized knowledge bases face a dilemma: merge everything into a single KB (losing domain specialization and polluting LLM context with irrelevant information) or require users to manually select the target KB. To the best of our knowledge, no existing Graph-RAG system discovers implicit structural connections between separate knowledge graphs at query time.

\paragraph{Contributions.} We present No\=esis, a fully implemented and validated distributed knowledge system that addresses these limitations through:

\begin{enumerate}
    \item A bidirectional graph traversal algorithm with forward-pass Graph-Feedback context resolution (n-gram scoring + recency decay) and backward-pass degree-based reconnection (\S\ref{sec:bidirectional});
    \item An AIMD concurrency controller that transfers TCP congestion control principles to RAG pipeline orchestration with Redis-backed persistent state (\S\ref{sec:aimd});
    \item \emph{Mo\=esis}: automatic domain-aware profiling, classification, and selective quantization of Mixture-of-Experts models with runtime re-adaptation (\S\ref{sec:moesis});
    \item \emph{Mesh}: hierarchical fingerprint-based cross-KB routing with runtime structural discovery and adaptive Natural Break thresholds (\S\ref{sec:mesh}).
\end{enumerate}

\section{System Architecture}
\label{sec:architecture}

No\=esis is a \emph{decoupled} architecture comprising five independent components communicating via HTTP REST and Redis message queues:

\begin{itemize}[leftmargin=*]
    \item \textbf{Cortex}: LLM inference engine (llama.cpp + Mo\=esis optimization). Runs on dedicated GPU hardware, independently scalable.
    \item \textbf{Neuron}: Knowledge graph extraction library. Implements bidirectional traversal, entity deduplication, and cross-linking.
    \item \textbf{Synapsis}: Orchestration layer with distributed job queue, AIMD controller, and API gateway.
    \item \textbf{Retina}: Audio/video ingestion module (C++ VAD engine + speech-to-text).
    \item \textbf{Mesh}: Cross-KB semantic router operating as a shared in-process kernel ($<2$\,ms latency).
\end{itemize}

\begin{figure}[t]
    \centering
    \includegraphics[width=\linewidth]{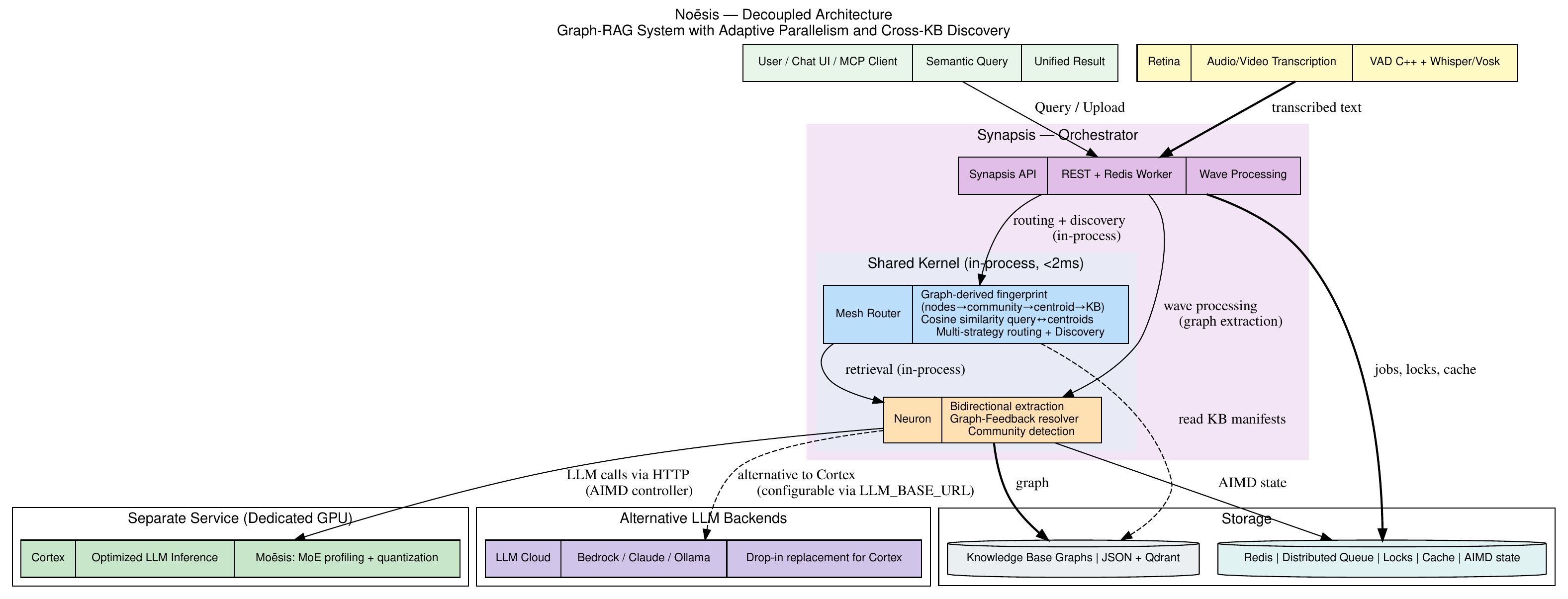}
    \caption{No\=esis system architecture. Five decoupled components communicate via REST APIs and Redis message queues, enabling independent scaling and backend replacement.}
    \label{fig:architecture}
\end{figure}

This separation is architecturally critical: the inference backend can be replaced (e.g., with AWS Bedrock, Claude API, or Ollama) without modifying the extraction or routing logic. The system adapts automatically to available hardware with \emph{zero manual configuration}, scaling from consumer GPUs to multi-GPU servers. Stress tests confirm stable operation even at 6\,GB VRAM with no crashes or data loss.

The extraction pipeline is \emph{source-agnostic}: Neuron operates on any textual corpus, including PDF and DOCX documents, web pages, audio/video transcriptions, and software source code repositories. This generality enables domain-specific knowledge graphs from heterogeneous input sources without pipeline modifications.

\section{Bidirectional Graph Traversal}
\label{sec:bidirectional}

\subsection{Problem Statement}

When processing a long document (e.g., 200+ pages), existing systems split it into fixed-size chunks and analyze each independently. A concept introduced in Chapter~1 and referenced in Chapter~10 produces two disconnected nodes instead of a single entity with cross-document edges. Rolling summary approaches (maintaining an LLM-generated summary between chunks) add significant overhead per long document and still miss long-range dependencies. In our tests, No\=esis maintains semantic continuity across single documents approaching 200 pages.

\begin{figure}[t]
    \centering
    \includegraphics[width=\linewidth]{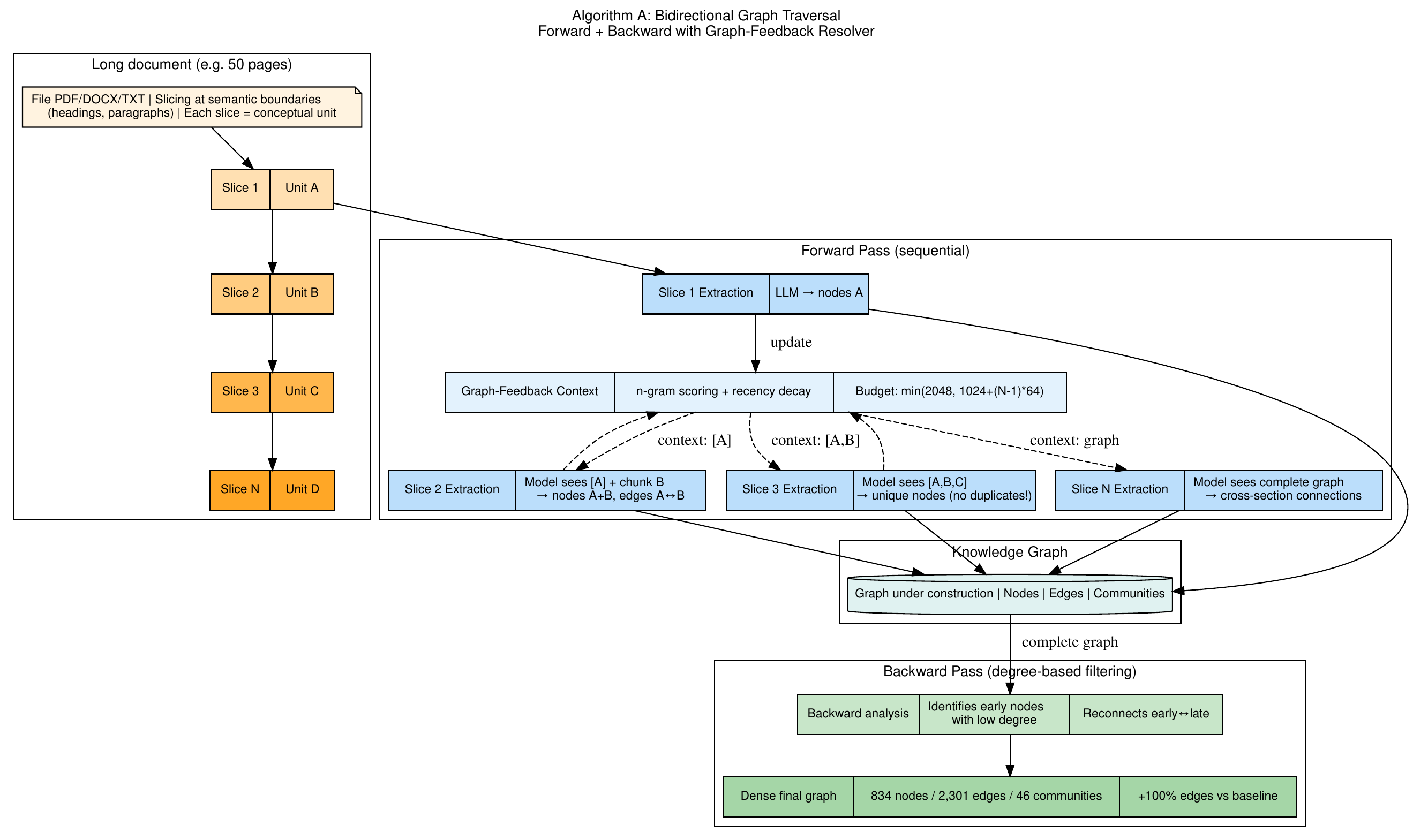}
    \caption{Bidirectional graph traversal: the forward pass builds the graph with degrading memory context, then the backward pass reconnects early low-degree nodes to semantically related late nodes.}
    \label{fig:bidirectional}
\end{figure}

\subsection{Forward Pass with Graph-Feedback Context}

No\=esis processes document slices sequentially, simulating human reading with \emph{degrading memory}. After each slice, the \textbf{Graph-Feedback Context Resolver} selects the most relevant previously-extracted nodes to provide as context for the next slice:

\begin{equation}
\text{score}(n) = \text{relevance}(n) + \text{recency}(n)
\end{equation}

where $\text{relevance}(n)$ is computed via unigram + bigram keyword overlap between node $n$'s label and the upcoming slice, and $\text{recency}(n)$ provides a bounded bonus for recently-extracted nodes that decays with document distance. This mimics human working memory: recent concepts are vivid; distant ones fade unless directly relevant.

The context budget grows adaptively with document length, bounded between a minimum (to avoid starving short documents) and a maximum (to prevent context explosion), scaling linearly with the number of slices.

Document slicing uses \emph{semantic boundaries} (headings, paragraph breaks) rather than arbitrary token offsets, ensuring each slice contains a complete conceptual unit.

\subsection{Backward Pass: Degree-Based Reconnection}

After the forward pass completes, a second pass identifies ``early'' nodes with low degree (few connections) and attempts to reconnect them to semantically related ``late'' nodes. This closes long-range dependencies that the forward pass could not see---analogous to a reader who, upon finishing a book, recognizes that an early character connects to a late plot event.

The backward pass activates only when the document is sufficiently long and produces enough nodes to benefit from reconnection; below these thresholds the graph is too small to benefit.

\subsection{Integrated Sub-Algorithms}

\paragraph{Entity Deduplication (4-phase pipeline).}
\begin{enumerate}
    \item \emph{Exact normalization} with per-source-file partitioning to avoid cross-document false merges.
    \item \emph{Locality-sensitive hashing (LSH) blocking + Jaro-Winkler} scoring with two innovations: (a)~a \textbf{Shannon entropy gate} that protects low-information nodes (e.g., ``Results'', ``Method'') from automatic merging by detecting insufficient label complexity; (b)~a \textbf{Leiden community boost} that increases the merge score when both nodes share a graph community.
    \item \emph{Variant pair detection}: identifies siblings differing only by numeric suffix (e.g., ASR1603 vs.\ ASR1604) and protects them from merge.
    \item \emph{LLM tiebreaker}: ambiguous pairs in an intermediate confidence band are resolved in batch via binary yes/no queries.
\end{enumerate}

\paragraph{Cross-Linking: Pistol/Cannon Strategy.}
For small graphs, cosine similarity of node embeddings suffices (``pistol''). For larger graphs, an additional LLM pass examines an \emph{ambiguity band} where embeddings are unreliable but the LLM can discover non-obvious causal relations (``cannon''). Example: discovering ``stress'' $\to$ ``magnesium deficiency'' $\to$ ``guided exercise''.

\paragraph{Adaptive Retry with Recursive Bisection.}
When the LLM produces truncated output, context overflow, or hollow responses, the system bisects the slice at a semantic boundary and retries recursively. This guarantees that no content is silently discarded: problematic slices are recursively bisected until successful extraction is achieved.

\paragraph{Prompt Injection Protection.}
Since No\=esis processes user-uploaded documents (PDF, DOCX), a malicious file could inject instructions into the LLM during extraction. Each source file is wrapped in cryptographically-tagged containment delimiters that mark content as untrusted, and known LLM control sequences are neutralized before extraction to prevent instruction injection.

\subsection{Measured Results}

On a multilingual knowledge base (5 PDF documents):
\begin{itemize}
    \item Baseline (single-pass, no feedback): 29 nodes, 36 edges, 4 communities.
    \item With bidirectional traversal: \textbf{253 nodes, 257 edges, 179 cross-links, 12 communities}.
    \item Full extraction run: 834 nodes, 2{,}301 edges, 46 communities.
    \item Backward pass contribution: substantial increase in cross-section edges versus forward-only baseline (magnitude varies by corpus structure).
\end{itemize}

\paragraph{Long-Range Reconnection Example.}
On a 193-page psychology book (\emph{Waking the Tiger: Healing Trauma}), the backward pass produces 928 intra-document causal edges. These are connections between concepts introduced in early chapters and therapeutic outcomes discussed hundreds of pages later---within a \emph{single} document. For example, \emph{Dissociation} (a defense mechanism defined in early chapters) is connected via an \texttt{inhibits} edge to \emph{Active Mobilization} (a therapeutic strategy introduced in later chapters). Similarly, \emph{Social Isolation} is linked to \emph{Trauma Resolution} via an inhibitory relationship spanning the full length of the text.

The Graph-Feedback Context Resolver makes these connections possible by providing relevant previously-extracted nodes as context during each slice's extraction. When processing a late chapter that discusses therapeutic mobilization, the resolver surfaces the \emph{Dissociation} node (extracted hundreds of pages earlier) as context---enabling the LLM to recognize and emit the causal relationship between them. This simulates a reader who, while reading Chapter~15, recalls a concept from Chapter~2 because it is relevant to the current content. Without this mechanism, each slice is processed in isolation and long-range relationships remain undiscovered---a limitation recently identified by CrossAug~\cite{crossaug2026} as a structural gap in chunk-independent extraction.

The resulting knowledge graph for this single document contains 299 nodes and 591 edges of five primary relational types (\texttt{enables}, \texttt{inhibits}, \texttt{leads\_to}, \texttt{resolves}, \texttt{transforms}) out of the 928 total causal edges---a density of structural knowledge that enables multi-hop therapeutic pathway queries across the full span of the text. An LLM-judged evaluation (GPT-4o, temperature~0) of 50 randomly sampled causal edges yields 72\% precision when the evaluator lacks book context, rising to 84\% when provided with domain context, and to 90\% upon source text verification---where each contested edge is checked against the original text. The five genuinely incorrect edges (10\%) consist of two inverted causal directions and three unsupported relationships. A second evaluation on a structurally different document---a clinical nutrition protocol in Russian (non-narrative, list-based format)---yields 48\% without context, rising to 74\% upon domain-contextualized re-evaluation. The consistent gap across both documents (18--26 points), across two domains, two languages, and two document structures, confirms that No\=esis extracts relationships so specific to the source material that external evaluators without document-level context cannot validate them. The 18-point gap between context-free evaluation (72\%) and source-verified precision (90\%), measured on this single-document sample, quantifies the value of the Graph-Feedback mechanism for domain-specific texts with specialized terminology. In chunk-independent extraction, the LLM processing a later chapter has no visibility into concepts introduced earlier; it therefore cannot emit cross-section causal relationships regardless of model capability.

\begin{figure}[t]
    \centering
    \includegraphics[width=0.75\linewidth]{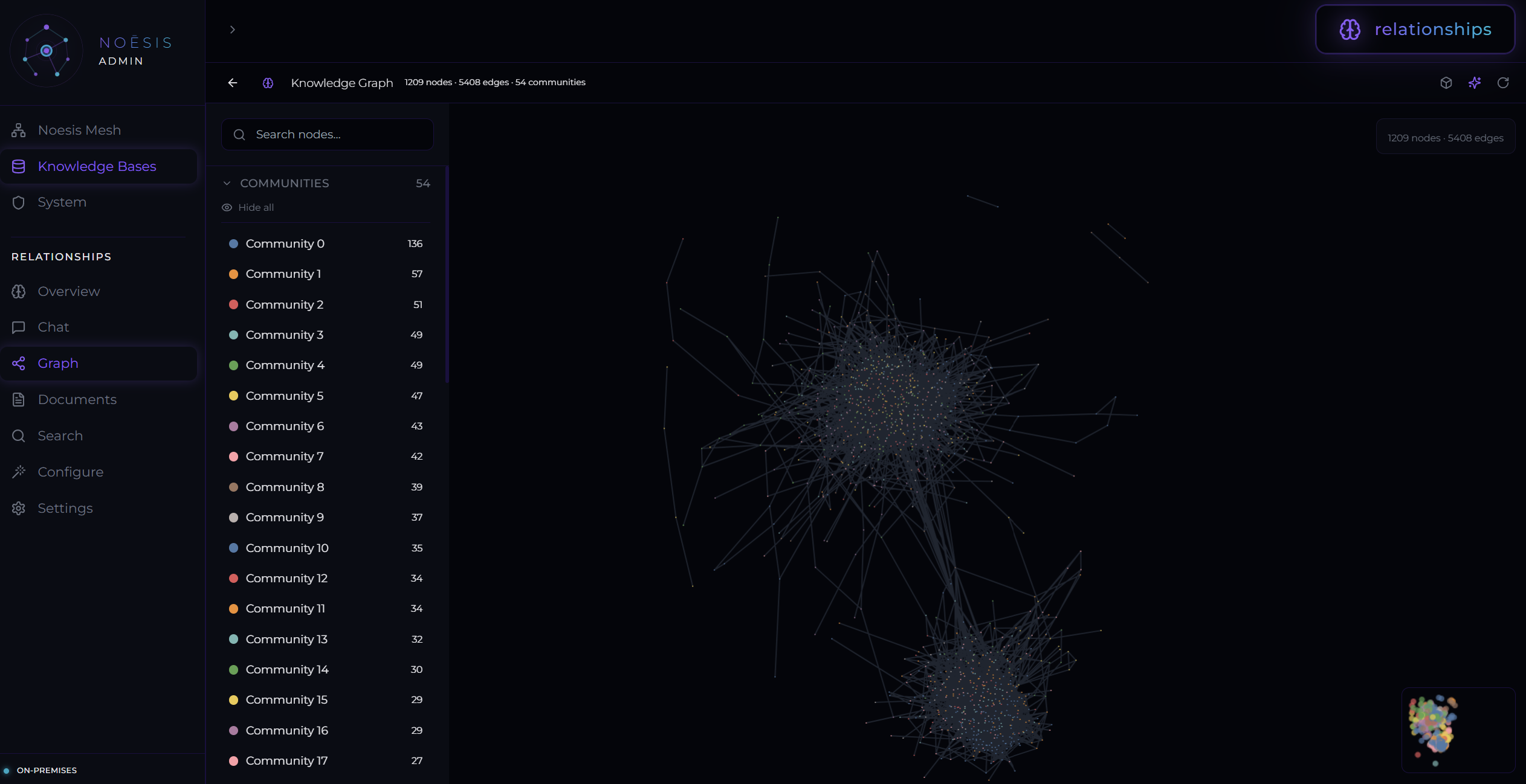}
    \caption{Intra-document cross-linking enabled by bidirectional traversal. The dense interconnection pattern between concepts from different document sections is a direct result of the backward pass reconnection. Single-pass extraction produces sparse, fragmented subgraphs without these long-range edges.}
    \label{fig:crosslinks}
\end{figure}

\section{AIMD Adaptive Parallel Processing}
\label{sec:aimd}

\subsection{The Hybrid Parallelism Problem}

Our bidirectional traversal requires intra-document sequentiality: slices within each document must be processed in order to build the Graph-Feedback context that the backward pass depends on. This constraint is fundamental to the algorithm and cannot be relaxed without losing the long-range edges that define its value.

Yet we still want to accelerate ingestion by processing multiple documents concurrently. The challenge is to maximize inter-document throughput without knowing in advance how much load the backend can handle---the same pipeline must operate on an RTX 4080 laptop, a cloud Bedrock endpoint, or an Ollama server, each with different capacity characteristics.

This creates a problem that existing concurrency control does not address: how to adapt document-level parallelism at runtime under multiple simultaneous constraints---intra-document sequentiality that forbids slice-level parallelism, inference backends with completely unknown capacity and failure characteristics (not just VRAM but latency profiles, rate limits, and transient error modes), crash recovery without warmup periods, and the requirement to remain fully agnostic to the backend type---while maintaining persistent state across distributed workers.

\subsection{Algorithm Design}

We apply the Additive Increase / Multiplicative Decrease (AIMD) principle---originally designed for TCP congestion control---to distributed RAG pipeline orchestration. The transfer requires adapting AIMD from its native context (single-connection, packet-level round-trip times) to a fundamentally different operating environment: a multi-worker job queue where (1)~concurrency is measured in documents, not packets; (2)~state must persist across worker crashes and restarts via a shared store (Redis); and (3)~the hybrid parallelism constraint (inter-document parallel, intra-document sequential) means that scaling decisions affect throughput differently than in unconstrained batch serving.
\begin{itemize}
    \item \textbf{Initial concurrency}: conservative start value.
    \item \textbf{Additive increase}: after a configurable streak of consecutive successes, $C \leftarrow C + 1$ (bounded by a system-defined maximum).
    \item \textbf{Multiplicative decrease}: after a configurable streak of consecutive failures, $C \leftarrow \lfloor C/2 \rfloor$.
\end{itemize}

The controller is \emph{backend-agnostic}: it adapts identically whether the inference backend is a local GPU (Cortex), a cloud API (AWS Bedrock, Claude), or a community server (Ollama). This is possible because No\=esis decouples inference from orchestration.

\subsection{Hybrid Parallelism}

A critical design choice: the system maintains \emph{inter-document parallelism} (multiple documents processed simultaneously) while preserving \emph{intra-document sequentiality} (slices within each document are processed in order). This is essential because the bidirectional traversal's degrading memory requires sequential slice processing to build correct cross-section connections.

\subsection{Measured Results}

\begin{itemize}
    \item 3 PDF documents, 13.4\,MB total: \textbf{1\,min\,6\,s} with adaptive parallelism vs.\ $\sim$25\,min sequential (\textbf{$23\times$ speedup}).
    \item Zero OOM events observed across all test configurations.
    \item Controller state survives worker crashes and restarts---no warmup period needed.
\end{itemize}

\section{Mo\=esis: Domain-Aware Selective Quantization}
\label{sec:moesis}

\subsection{Problem Statement}

Modern Mixture-of-Experts (MoE) models (e.g., Qwen3.6-35B-A3B with 256 experts per layer) load all experts into GPU memory even though only $\sim$8 are activated per token. On consumer hardware (6--12\,GB VRAM), this causes immediate OOM. Several recent approaches address mixed-precision allocation for MoE models: DynaExq~\cite{dynaexq2025} implements runtime dynamic expert-level quantization with hotness profiling and precision transitions; MoPEQ~\cite{mopeq2025} assigns optimal bit-width per expert using Hessian trace approximation; APEX~\cite{apex2026} performs per-tensor, per-layer precision allocation based on architectural role; and Mixture-Compressor~\cite{mixturecompressor2024} folds expert activation frequency into per-expert bit-width allocation. Mo\=esis differs from all of these in its end-to-end integration of domain-specific intelligence into the quantization lifecycle: rather than relying on architecture-based heuristics or generic calibration data, it derives compression decisions from actual expert activation patterns observed on a user-provided domain-representative sample, applies a promote-only GPU placement strategy that eliminates repeated CPU$\leftrightarrow$GPU data transfers after selective quantization, and preserves full-precision weights as a reference for runtime re-adaptation when the deployment domain changes---preventing cumulative precision loss across adaptation cycles.

\subsection{Three-Level Pipeline}

\begin{figure}[t]
    \centering
    \includegraphics[width=\linewidth]{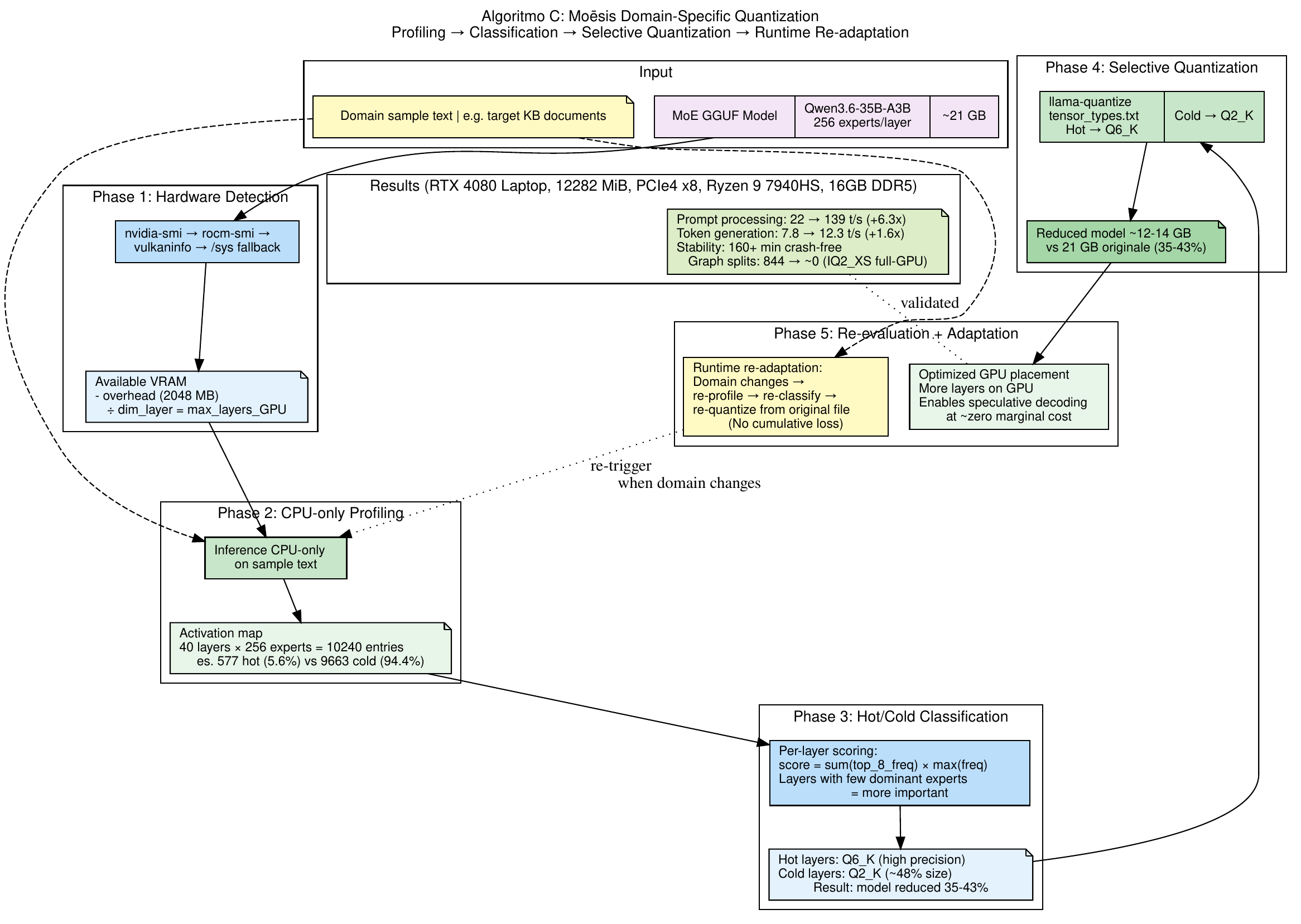}
    \caption{Mo\=esis three-level pipeline: (1)~domain-aware profiling identifies hot/cold experts, (2)~selective quantization applies different precision per layer, (3)~runtime re-adaptation re-profiles when the domain changes.}
    \label{fig:moesis}
\end{figure}

\paragraph{Level 1: Automatic Domain Profiling.}
Cortex runs CPU-only inference on a domain-representative text sample, recording per-layer per-expert activation frequencies. The resulting activation map captures which experts are relevant to the target domain. Typical outcome: only $\sim$5\% of experts are ``hot'' (frequently activated for this domain); $\sim$95\% are ``cold''.

\paragraph{Level 2: Selective Quantization and GPU Promotion.}
Each layer is classified using a scoring function that captures both the concentration of expert activations and the overall frequency of the most-used experts for the target domain. Hot layers (high concentration) are quantized at high precision (6-bit); cold layers at aggressive compression (2-bit, $\sim$48\% size reduction). The model shrinks from 21\,GB to $\sim$16\,GB on disk ($\sim$24\% reduction).

The reduced model size enables a \emph{promote-only} placement strategy: after quantization, the system re-evaluates how many layers fit entirely in GPU memory. Layers are promoted from CPU to GPU---never demoted---until VRAM is filled. This eliminates or substantially reduces the repeated CPU$\leftrightarrow$GPU data transfers that dominate inference latency on VRAM-constrained hardware. The profiling, classification, quantization, and placement steps execute automatically without manual configuration.

\paragraph{Level 3: Runtime Re-Adaptation.}
When the domain changes (e.g., from medical documents to engineering), the system re-executes profiling and re-quantizes from the \emph{original full-precision model} (preserved as backup at first quantization). This prevents cumulative precision loss across adaptation cycles---a critical distinction from static quantization approaches.

\subsection{Measured Results}

Table~\ref{tab:moesis} reports Mo\=esis performance on consumer hardware.

\begin{table}[H]
\centering
\caption{Mo\=esis performance on consumer hardware. The 6\,GB configuration represents a laboratory stress test demonstrating system stability under extreme resource constraints, not a recommended operational configuration.}
\label{tab:moesis}
\begin{tabular}{lcc}
\toprule
\textbf{Metric} & \textbf{RTX 4080 Laptop (12\,GB)} & \textbf{RTX 3060 (6\,GB) [stress test]} \\
\midrule
Prompt processing & $22 \to 139$ t/s ($6.3\times$) & Stable (2--6 t/s) \\
Token generation & $7.8 \to 12.3$ t/s ($1.6\times$) & Stable (2--6 t/s) \\
Stability & Continuous & 160+ min, zero crashes \\
Without Mo\=esis & Functional but slow & \textbf{Immediate OOM} \\
\bottomrule
\end{tabular}
\end{table}

Mo\=esis is not merely a speed optimization---it is an \emph{enabler}: on 6\,GB VRAM, the system does not start without it. The 6\,GB configuration demonstrates architectural resilience (zero crashes over 160+ minutes of continuous operation), while higher-VRAM configurations achieve substantially faster processing as more layers are promoted to GPU.

\section{Mesh: Cross-KB Routing and Discovery}
\label{sec:mesh}

\subsection{Problem Statement}

Organizations maintaining multiple specialized knowledge bases face a fundamental tension. A single monolithic KB (``KB-GOD'') pollutes the LLM context with irrelevant cross-domain information, degrading response quality. Separate KBs require users to know which domain to query and cannot discover implicit connections between domains. To the best of our knowledge, no existing Graph-RAG system implements automatic cross-KB routing with runtime structural discovery.

\subsection{Hierarchical Fingerprint Routing}

\begin{figure}[t]
    \centering
    \includegraphics[width=\linewidth]{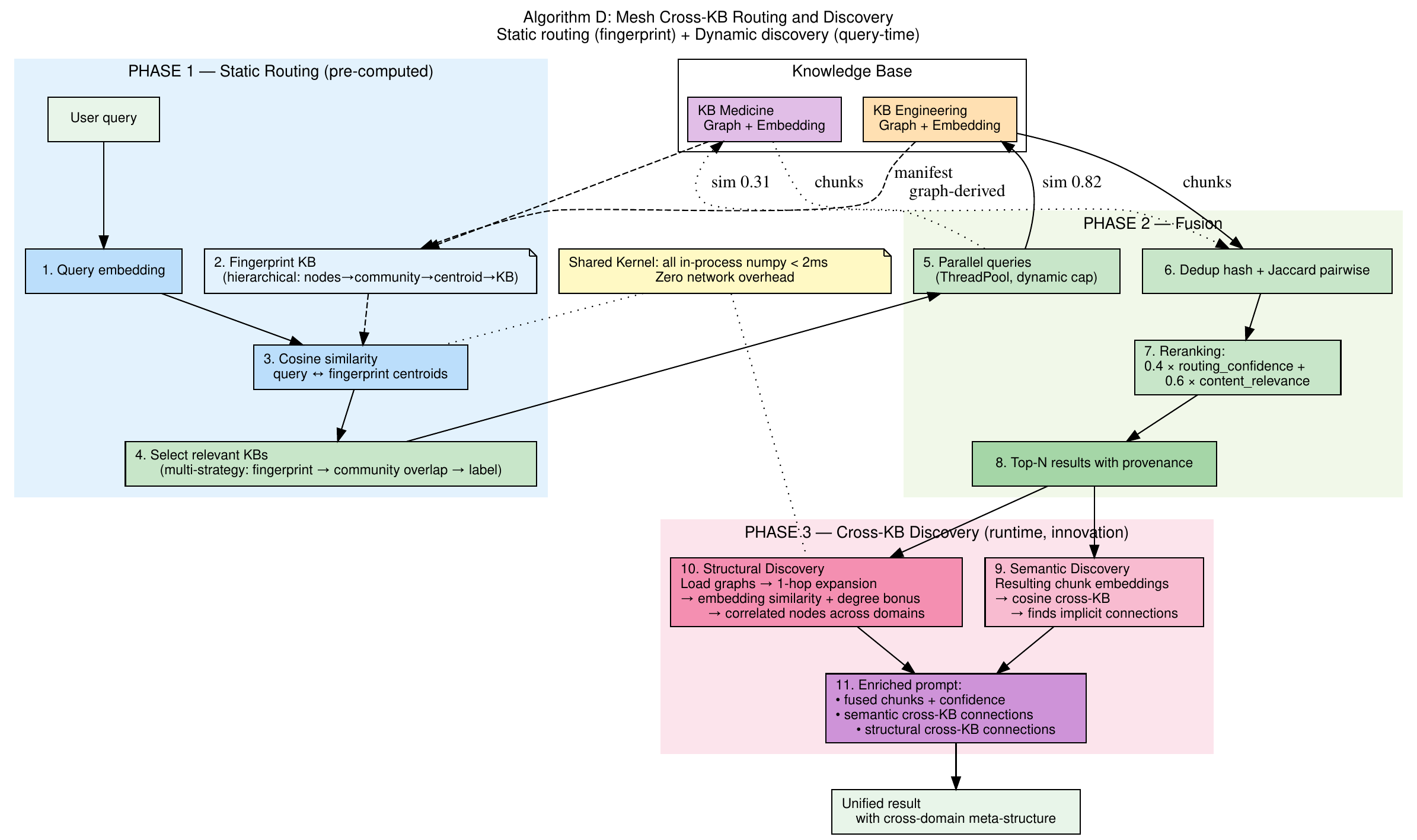}
    \caption{Mesh cross-KB routing: queries are routed to relevant knowledge bases via hierarchical fingerprint matching, then runtime structural discovery identifies emergent connections across separate graphs.}
    \label{fig:mesh}
\end{figure}

Each KB exports a semantic fingerprint derived from its graph structure:
\begin{equation}
\mathbf{fp}_{\text{KB}} = \text{L2-norm}\left(\sum_{c \in \mathcal{C}} w_c \cdot \boldsymbol{\mu}_c\right)
\end{equation}
where $\mathcal{C}$ is the set of Leiden communities, $w_c$ is the community weight, and $\boldsymbol{\mu}_c$ is the mean embedding of community $c$'s member nodes.

Query routing uses a multi-strategy approach with fallback:
\begin{enumerate}
    \item \emph{Primary}: cosine similarity between query embedding and KB fingerprints.
    \item \emph{Secondary}: community-level semantic overlap.
    \item \emph{Tertiary}: lexical label matching (ensures routing even without an active embedding model).
\end{enumerate}

Selected KBs are queried in parallel via a thread pool. Results undergo 5-phase fusion: extraction $\to$ confidence scoring (weighted combination of routing affinity and content relevance) $\to$ Jaccard n-gram deduplication $\to$ reranking $\to$ top-k cap.

\subsection{Runtime Structural Discovery}

After initial retrieval, Mesh discovers emergent connections between KBs through two mechanisms:

\paragraph{Chunk-Level Semantic Discovery.}
Computes embeddings of retrieved chunks and identifies cross-KB pairs with high similarity that were not pre-computed in any manifest.

\paragraph{Node-Level Structural Discovery.}
Loads graphs of selected KBs, identifies query-relevant nodes, expands to 1-hop neighbors, and compares nodes across KBs using embedding similarity with a structural bonus based on graph role similarity (incorporating node degree and community membership).

\subsection{Natural Break Adaptive Threshold}

Instead of a fixed global similarity threshold, Mesh computes the optimal threshold for each specific query:

\begin{enumerate}
    \item Compute all pairwise similarities between chunks/nodes of different KBs.
    \item Sort in descending order and compute first differences (slopes).
    \item Detect the ``knee'': the index where the slope exceeds a statistically-derived threshold based on the distribution's mean and standard deviation.
    \item Fallback: if no clear knee exists, use a high percentile as threshold.
    \item Safety clamp: the adaptive threshold never falls below a fixed domain-calibrated baseline.
\end{enumerate}

\paragraph{Behavior.} For semantically close KBs (e.g., two domains sharing overlapping terminology), the threshold lowers automatically, discovering real connections that a fixed threshold would filter. For distant KBs (e.g., domains with no shared vocabulary), it remains high, filtering noise.

\subsection{Measured Results}

\paragraph{Evaluation example.}
Query: ``How can stress create hormonal dysfunctions and affect relationships? What methods mitigate negative effects?''\\
KBs involved: three domain-specific knowledge bases covering health, behavioral science, and interpersonal dynamics.\\
Model: Gemma~4 E2B (2.3B effective parameters, designed for smartphones, $<$2\,GB RAM).

\begin{itemize}
    \item \textbf{Without} Mesh discovery: Fragmented response with explicit disclaimers about incomplete information. Model admits inability to connect domains.
    \item \textbf{With} Mesh discovery + Natural Break: Organic, confident response connecting cortisol $\to$ endocrine system $\to$ estrogen/progesterone $\to$ emotional regulation $\to$ couple communication. Proposes structured strategies (individual: mindfulness; couple: empathic listening, physical connection $\to$ oxytocin). Zero disclaimers.
\end{itemize}

The emergent connection ``physical contact $\to$ oxytocin $\to$ couple bonding'' exists in no single document---it was discovered by Mesh at query time by comparing nodes across separate graphs. A 2.3B-parameter model achieves multi-hop cross-domain reasoning that the same model cannot perform without Mesh---demonstrating that the architecture, not model scale, enables emergent discovery.

Critically, Mesh is equally effective at \emph{rejecting} irrelevant connections: a knowledge base built from a software codebase correctly produces zero cross-KB connections with semantically unrelated domains, demonstrating that the routing is selective rather than indiscriminate---avoiding the context pollution that degrades monolithic approaches.

\paragraph{Shared Kernel Architecture.}
Routing and discovery execute \emph{within} the API process itself (in-process NumPy operations), not as external service calls. This Shared Kernel pattern eliminates network round-trips entirely, achieving routing latency $<$2\,ms. The same in-process execution applies to runtime structural discovery---enabling real-time cross-KB reasoning without the latency penalty of microservice architectures.

\section{Evaluation}
\label{sec:evaluation}

Table~\ref{tab:comparison} summarizes the capabilities introduced by No\=esis relative to common design patterns in existing Graph-RAG systems.

\begin{table}[H]
\centering
\small
\caption{Capabilities of No\=esis versus common patterns in existing Graph-RAG approaches$^{\dagger}$.}
\label{tab:comparison}
\begin{tabular}{p{4.2cm}p{4.8cm}c}
\toprule
\textbf{Capability} & \textbf{Existing Approaches} & \textbf{No\=esis} \\
\midrule
Cross-chunk context during extraction & Rare or limited & \checkmark \\
Adaptive ingestion parallelism & Not observed & \checkmark \\
Independently scalable inference backend & Configurable backends, not decoupled services & \checkmark \\
MoE-aware selective quantization & Architecture-based or generic calibration; static & \checkmark \\
Cross-KB routing ($<2$\,ms) & Not addressed & \checkmark \\
Cross-KB structural discovery (runtime) & Not addressed & \checkmark \\
Adaptive similarity thresholds & Fixed thresholds & \checkmark \\
Multi-pass entity deduplication & Basic or none & \checkmark \\
Extraction-time prompt injection containment & Not addressed & \checkmark \\
6\,GB stress test $\to$ multi-GPU scaling & Typically requires $>$16\,GB & \checkmark \\
\bottomrule
\end{tabular}
\end{table}

\vspace{0.5em}
\noindent$^{\dagger}$\small\emph{Reflects common patterns observed in published documentation of systems including GraphRAG, LightRAG, LazyGraphRAG, HippoRAG, DynaExq, MoPEQ, APEX, and RAGFlow as of August 2026. Individual systems may address some capabilities through extensions or plugins not reflected here.}

\begin{figure}[t]
    \centering
    \includegraphics[width=0.75\linewidth]{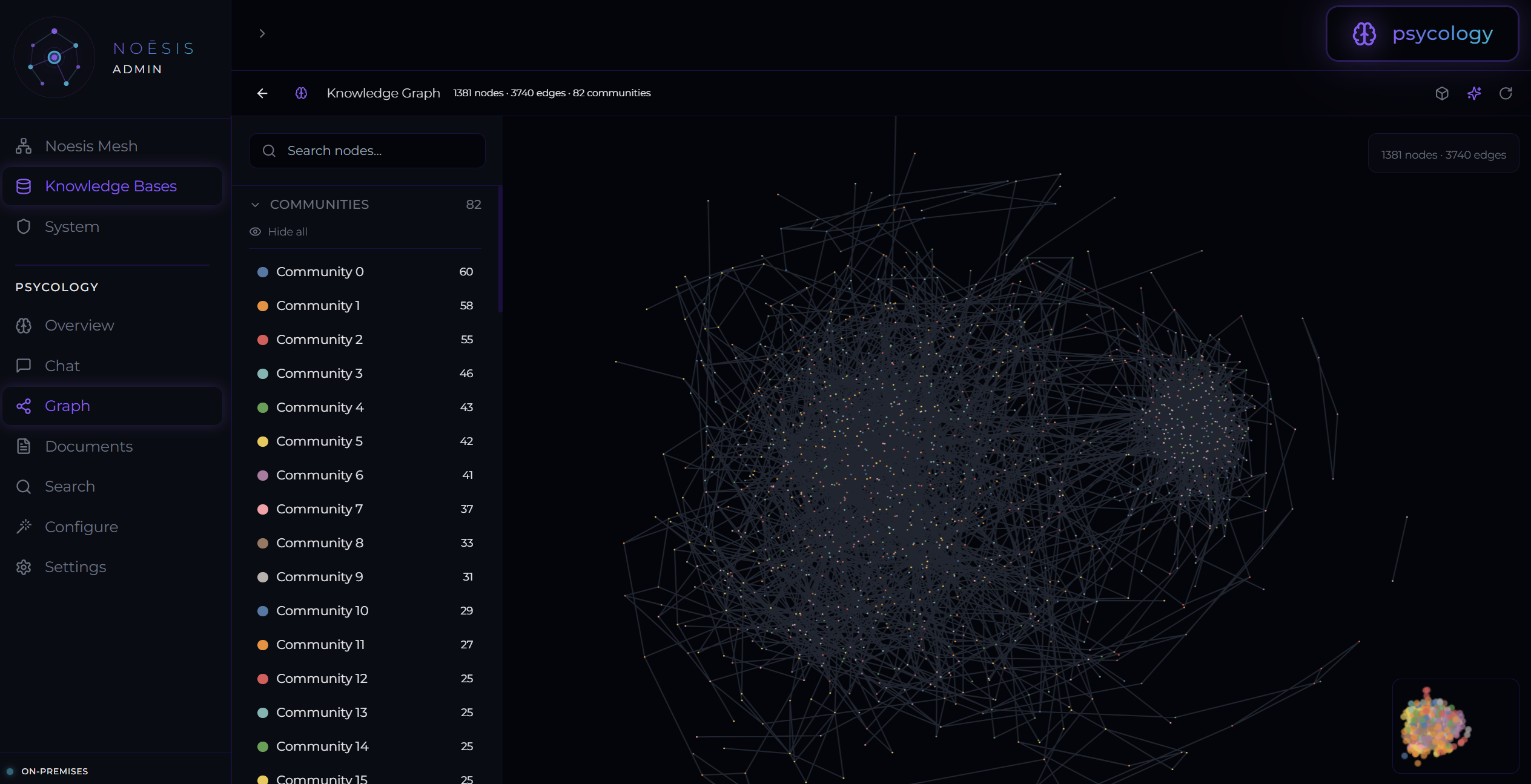}
    \caption{Two-dimensional visualization of a knowledge graph generated by No\=esis (1{,}381 nodes, 3{,}740 edges, 82 communities). Node colors indicate Leiden communities; edge thickness reflects relationship strength.}
    \label{fig:graph3d}
\end{figure}

\subsection{Multi-Hop QA Benchmark}
\label{sec:eval-hotpotqa}

To evaluate retrieval quality on an established multi-hop reasoning task, we benchmark No\=esis on the HotpotQA validation set (distractor setting)~\cite{yang2018hotpotqa}, following the same protocol and subset size (1{,}000 questions) used by HippoRAG~\cite{gutierrez2024hipporag} and StepChain~\cite{li2025stepchain}.

\paragraph{Setup.}
The No\=esis knowledge graph is constructed by Qwen3.6-35B-A3B running on-premises using domain-appropriate extraction prompts with no task-specific optimization or training. Answer generation is performed by GPT-4o. Baseline systems---GraphRAG, HopRAG, and StepChain---use GPT-4o for \emph{both} graph construction and answer generation, representing a substantially more expensive configuration for the extraction phase. BGE dense retrieval uses bge-large-en-v1.5 embeddings with GPT-4o for answer generation. Additionally, No\=esis retrieves $k{=}10$ chunks per query, whereas baseline systems in~\cite{li2025stepchain} use $k{=}20$ passages---a retrieval budget disadvantage of 50\% for our system that makes the achieved scores more notable.

\paragraph{Results.}
Table~\ref{tab:hotpotqa} reports Exact Match (EM) and F1 scores. All baseline numbers are taken from the StepChain paper~\cite{li2025stepchain}.

\begin{table}[H]
\centering
\caption{Multi-hop QA performance on HotpotQA (distractor, 1{,}000 questions). Baselines from~\cite{li2025stepchain}. $\dagger$~denotes systems using GPT-4o for both graph construction and answer generation.}
\label{tab:hotpotqa}
\begin{tabular}{lcc}
\toprule
\textbf{System} & \textbf{EM} & \textbf{F1} \\
\midrule
GraphRAG$^{\dagger}$~\cite{edge2024graphrag} & 31.70 & 42.74 \\
BGE dense + GPT-4o & 47.60 & 60.36 \\
\textbf{No\=esis} + GPT-4o & \textbf{59.50} & \textbf{74.74} \\
HopRAG$^{\dagger}$~\cite{hoprag2025} & 62.00 & 76.06 \\
StepChain$^{\dagger}$~\cite{li2025stepchain} & 66.70 & 79.50 \\
\bottomrule
\end{tabular}
\end{table}

No\=esis achieves 59.50 EM / 74.74 F1, surpassing both GraphRAG (+27.8~EM) and dense retrieval (+11.9~EM), while using a 35B on-premises model for the computationally expensive graph construction phase rather than GPT-4o. Notably, HotpotQA paragraphs average $\sim$4 sentences---below the threshold at which No\=esis's bidirectional traversal and backward pass activate. The system's primary architectural advantage (long-range cross-section reconnection) is structurally inactive on this benchmark, yet it still achieves 96\% of HopRAG's performance. The remaining gap to StepChain ($-$7.2~EM) and HopRAG ($-$2.5~EM) reflects primarily the difference in graph construction model capacity (35B on-premises vs.\ GPT-4o).

\paragraph{Ablation: Architecture vs.\ Model Scale.}
To isolate the contribution of the retrieval architecture from the answer-generation LLM, we replace GPT-4o with Gemma~4 E2B (2.3B effective parameters, designed for smartphones, $<$2\,GB RAM). Results are shown in Table~\ref{tab:ablation}.

\begin{table}[H]
\centering
\caption{Ablation: effect of answer-generation model on HotpotQA. The No\=esis graph (built by Qwen3.6-35B-A3B) is identical across both configurations.}
\label{tab:ablation}
\begin{tabular}{lcc}
\toprule
\textbf{Configuration} & \textbf{EM} & \textbf{F1} \\
\midrule
BGE dense + GPT-4o & 47.60 & 60.36 \\
No\=esis + Gemma~4 E2B (2.3B) & 47.20 & 60.30 \\
No\=esis + GPT-4o & 59.50 & 74.74 \\
\bottomrule
\end{tabular}
\end{table}

No\=esis with a 2.3B-parameter answer model (EM=47.20, F1=60.30) matches the performance of BGE dense retrieval with GPT-4o (EM=47.60, F1=60.36)---a model approximately $100\times$ larger. This suggests that the retrieval architecture contributes approximately 80\% of the overall QA performance, with LLM reasoning capability accounting for the remaining $\sim$20\%. The graph structure delivers sufficiently precise context that even a small model can produce correct answers.

\paragraph{Ablation: Retrieval Budget.}
To verify whether the $k{=}10$ retrieval budget disadvantages No\=esis relative to baselines operating at $k{=}20$, we re-run the full 1{,}000-question evaluation with $k{=}20$ chunks. Results: EM=60.90, F1=76.12. Doubling the retrieval budget yields only $+1.4$~EM, indicating that graph-guided retrieval already captures the relevant multi-hop evidence at lower~$k$. At matched budget ($k{=}20$), No\=esis matches HopRAG in F1 (76.12 vs.\ 76.06) while using a 35B on-premises model for graph construction rather than GPT-4o.

\subsection{Source Code Understanding}
\label{sec:eval-code}

To demonstrate the source-agnostic nature of the architecture, we evaluate No\=esis on a software source code corpus (55 Python files). The system produces a knowledge graph of 978 nodes and 3{,}824 edges with density comparable to document-based KBs of similar corpus size.

\paragraph{Query Evaluation.}
Query: \emph{``Trace the execution flow of a mesh chat query through the system.''}

The system correctly identifies the complete 8-step execution pipeline spanning 5+ modules, from HTTP endpoint through orchestration, parallel KB queries, result fusion, cross-KB discovery, to streaming response. This demonstrates that No\=esis delivers equivalent reasoning quality on source code as on natural language documents, without pipeline modifications---confirming the source-agnostic design claimed in \S\ref{sec:architecture}.

\section{Related Work}
\label{sec:related}

\textbf{Graph-RAG Systems.} Microsoft GraphRAG~\cite{edge2024graphrag} introduced community-based summarization but uses static chunking and sequential processing. LightRAG~\cite{guo2024lightrag} reduces indexing cost but does not address cross-document continuity. LazyGraphRAG~\cite{lazygraphrag2024} eliminates pre-summarization but still chunks independently. HippoRAG~\cite{gutierrez2024hipporag} models hippocampal memory for retrieval and extracts simplified triples (OpenIE), but does not build dense inter-document graph structures. CrossAug~\cite{crossaug2026} addresses the same underlying problem---missing cross-chunk relations---through a complementary approach: a GNN-guided post-extraction augmentation step that identifies high-scoring regions for LLM-based relation completion. Our bidirectional traversal differs by propagating previously-extracted graph structure as context \emph{during} the extraction process itself, combined with a backward reconnection pass that closes long-range dependencies.

\textbf{LLM Concurrency Control.} CONCUR~\cite{concur2026} applies AIMD-based admission control to regulate the number of active agents in LLM batch inference, operating at the GPU KV-cache level within a single serving engine. HiveMind~\cite{hivemind2026} uses AIMD backpressure as part of an HTTP proxy for coordinating concurrent LLM agent workloads. These systems operate on unconstrained batch serving where any request can be processed in parallel with any other.

Our controller addresses a fundamentally different problem. The bidirectional traversal algorithm requires intra-document sequentiality: slices within a single document must be processed in order to build the Graph-Feedback context that the backward pass depends on. This constraint is intrinsic to the extraction method and cannot be relaxed. The concurrency controller must therefore operate on \emph{document-level} granularity (not token, not KV-cache, not agent), adapt to backends with unknown capacity (consumer GPU, cloud API, or local Ollama---no prior specification required), maintain persistent state across distributed workers via a crash-resilient shared store, and resume operation without warmup after worker failure. To the best of our knowledge, no surveyed system combines these properties.

\textbf{MoE Optimization.} Recent work on Mixture-of-Experts quantization has explored multiple approaches to mixed-precision allocation. DynaExq~\cite{dynaexq2025} implements a runtime system with hotness-aware precision profiling, non-blocking precision transitions, and fragmentation-free memory pooling for memory-constrained GPU inference. MoPEQ~\cite{mopeq2025} assigns optimal bit-width per expert using Hessian trace approximation. APEX~\cite{apex2026} performs per-tensor, per-layer precision allocation based on architectural role and layer sensitivity. Mixture-Compressor~\cite{mixturecompressor2024} folds expert activation frequency into per-expert bit-width allocation. FIDDLER~\cite{fiddler2025} profiles expert popularity for CPU-GPU placement but does not compress the model. HybriMoE~\cite{hybrimoe2025} optimizes CPU-GPU scheduling on the kTransformers framework. ExpertFlow~\cite{expertflow2024} optimizes expert caching for inference without model compression. imatrix supports domain-specific calibration data but still quantizes all layers uniformly.

Mo\=esis differs from all of the above in its end-to-end integration of domain-specific intelligence into the quantization lifecycle: it derives compression decisions from actual expert activation patterns observed on a user-provided domain-representative sample, applies a promote-only GPU placement strategy that eliminates repeated CPU$\leftrightarrow$GPU transfers after selective quantization, and preserves full-precision weights as a reference for runtime re-adaptation when the deployment domain changes---preventing cumulative precision loss across adaptation cycles.

\textbf{Multi-KB Routing.} R1-Router~\cite{r1router2025} trains an LLM via reinforcement learning to decide when and where to retrieve from multiple KBs during step-wise reasoning. DAKS~\cite{daks2026} performs KB routing with budgeted retrieval and alignment graphs for cross-KB evidence fusion. HydraRAG~\cite{hydrarag2025} combines graph topology with tri-factor cross-source verification. Adaptive-k~\cite{adaptivek2025} uses largest-gap detection for retrieval quantity selection---a different problem from connection discovery thresholds. These systems route queries to relevant KBs but do not discover \emph{emergent structural connections} between separate knowledge graphs at query time. Mesh differs by performing runtime comparison of graph nodes and chunks across KBs using adaptive Natural Break thresholds---discovering connections that exist in no single KB's index.

\section{Limitations and Future Work}
\label{sec:limitations}

The measured speedup values ($23\times$ for AIMD parallelism, $6.3\times$ for Mo\=esis prompt processing) depend on the specific hardware and corpus tested. On different corpora or with different backend configurations, the absolute values will vary, although we expect the qualitative trends to hold.

The evaluation presented in this paper spans a standard multi-hop QA benchmark (HotpotQA), two long documents in different languages and formats, a software codebase, and three cross-domain knowledge bases. Broader evaluation across additional benchmarks and domains would further characterize the system's boundaries.

The Mesh cross-KB routing evaluation demonstrates the mechanism on a representative query across three domain-specific KBs. Characterizing performance across a wider range of query types and KB configurations remains ongoing work.

\section{Conclusion}
\label{sec:conclusion}

We have presented No\=esis, a fully implemented Graph-RAG system that introduces four algorithmic innovations addressing persistent limitations of existing approaches. The bidirectional traversal with Graph-Feedback context produces knowledge graphs with densities significantly exceeding those of independent chunking, achieving 90\% source-verified precision on long-range causal edges extracted from a 193-page document. The AIMD concurrency controller adapts document-level parallelism at runtime under the constraint of intra-document sequentiality---a requirement intrinsic to the bidirectional traversal algorithm---enabling safe, adaptive throughput across heterogeneous hardware without manual configuration. Mo\=esis makes MoE models more viable on consumer GPUs through domain-aware selective quantization with runtime re-adaptation. And Mesh enables automatic cross-domain reasoning without knowledge base fusion, achieving emergent multi-hop discovery on small on-premises models.

The system has been fully implemented and extensively tested on a corpus exceeding 60 documents and 170\,MB across multiple knowledge bases---including PDF, DOCX, a complete software codebase, and single documents approaching 200 pages---processing documents in multiple languages (including non-Latin scripts) and heterogeneous source types. Quantitative evaluation on HotpotQA (\S\ref{sec:eval-hotpotqa}) demonstrates competitive multi-hop retrieval quality, surpassing GraphRAG by +27.8~EM while using a 35B on-premises model for graph construction.

\bibliographystyle{plain}

\begin{thebibliography}{15}

\bibitem{edge2024graphrag}
D.~Edge, H.~Trinh, N.~Cheng, et al.
\newblock From Local to Global: A Graph RAG Approach to Query-Focused Summarization.
\newblock \emph{arXiv preprint arXiv:2404.16130}, 2024.

\bibitem{guo2024lightrag}
Z.~Guo, L.~Zhao, et al.
\newblock LightRAG: Simple and Fast Retrieval-Augmented Generation.
\newblock \emph{EMNLP}, 2025.

\bibitem{gutierrez2024hipporag}
B.~Gutierrez, Y.~Yang, et al.
\newblock HippoRAG: Neurobiologically Inspired Long-Term Memory for Large Language Models.
\newblock \emph{arXiv preprint arXiv:2405.14831}, 2024.

\bibitem{concur2026}
Q.~Chen et al.
\newblock CONCUR: High-Throughput Agentic Batch Inference of LLM via Congestion-Based Concurrency Control.
\newblock \emph{arXiv preprint arXiv:2601.22705}, 2026.

\bibitem{lazygraphrag2024}
Microsoft Research.
\newblock LazyGraphRAG: Setting a new standard for quality and cost.
\newblock \emph{Microsoft Research Blog}, 2024.
\newblock \url{https://www.microsoft.com/en-us/research/blog/lazygraphrag-setting-a-new-standard-for-quality-and-cost/}

\bibitem{adaptivek2025}
Adaptive-k Authors.
\newblock Efficient Context Selection for Long-Context QA.
\newblock \emph{arXiv preprint arXiv:2506.08479}, 2025.

\bibitem{dynaexq2025}
DynaExq Authors.
\newblock DynaExq: Dynamic Expert Quantization for Scalable Mixture-of-Experts Inference.
\newblock \emph{arXiv preprint arXiv:2511.15015}, 2025.

\bibitem{mopeq2025}
K.~T.~Chitty-Venkata, J.~Ye, M.~Emani.
\newblock MoPEQ: Mixture of Mixed Precision Quantized Experts.
\newblock \emph{Proceedings of the IEEE/CVF ICCV Workshops}, 2025.

\bibitem{apex2026}
E.~Di Giacinto, R.~Palethorpe.
\newblock APEX: Adaptive Precision for Expert Models.
\newblock Technical Report, LocalAI, March 2026.
\newblock \url{https://github.com/localai-org/apex-quant}

\bibitem{fiddler2025}
FIDDLER Authors.
\newblock FIDDLER: CPU-GPU Orchestration for Fast Inference of Mixture-of-Experts Models.
\newblock \emph{ICLR}, 2025.

\bibitem{hybrimoe2025}
HybriMoE Authors.
\newblock HybriMoE: Hybrid CPU-GPU Scheduling and Cache Management for Efficient MoE Inference.
\newblock \emph{arXiv preprint arXiv:2504.05897}, 2025.

\bibitem{expertflow2024}
ExpertFlow Authors.
\newblock ExpertFlow: Optimized Expert Activation and Token Allocation for Efficient Mixture-of-Experts Inference.
\newblock \emph{arXiv preprint arXiv:2410.17954}, 2024.

\bibitem{hivemind2026}
HiveMind Authors.
\newblock HiveMind: HTTP Proxy with AIMD Backpressure for Concurrent LLM Agent Workloads.
\newblock \emph{arXiv preprint arXiv:2604.17111}, 2026.

\bibitem{r1router2025}
C.~Peng, Z.~Xu, Z.~Liu, et al.
\newblock R1-Router: Learning to Route Queries Across Knowledge Bases for Step-wise Retrieval-Augmented Reasoning.
\newblock \emph{arXiv preprint arXiv:2505.22095v1}, 2025.

\bibitem{daks2026}
DAKS Authors.
\newblock Traceable Cross-Source RAG for Chinese Tibetan Medicine Question Answering.
\newblock \emph{arXiv preprint arXiv:2602.05195}, 2026.

\bibitem{hydrarag2025}
HydraRAG Authors.
\newblock HydraRAG: Structured Cross-Source Enhanced Large Language Model Reasoning.
\newblock \emph{arXiv preprint arXiv:2505.17464}, 2025.

\bibitem{crossaug2026}
CrossAug Authors.
\newblock CrossAug: GNN-Guided Cross-Chunk Graph Augmentation for Graph-RAG.
\newblock \emph{arXiv preprint arXiv:2605.28004}, 2026.

\bibitem{yang2018hotpotqa}
Z.~Yang, P.~Qi, S.~Zhang, Y.~Bengio, W.~Cohen, R.~Salakhutdinov, C.~Manning.
\newblock HotpotQA: A Dataset for Diverse, Explainable Multi-hop Question Answering.
\newblock \emph{EMNLP}, 2018.

\bibitem{li2025stepchain}
T.~Ni, X.~Yuan, W.~Zhang, S.~Li, K.~Wu, R.~P.~Liu, W.~Ni.
\newblock StepChain GraphRAG: Reasoning Over Knowledge Graphs for Multi-Hop Question Answering.
\newblock \emph{arXiv preprint arXiv:2510.02827}, 2025.

\bibitem{hoprag2025}
H.~Liu et al.
\newblock HopRAG: Multi-Hop Reasoning for Logic-Aware Retrieval-Augmented Generation.
\newblock \emph{ACL Findings}, 2025.

\bibitem{mixturecompressor2024}
Mixture-Compressor Authors.
\newblock Mixture-Compressor: Expert-Aware Quantization for Mixture-of-Experts Models.
\newblock \emph{arXiv preprint}, 2024.

\end{thebibliography}
{\footnotesize\setlength{\itemsep}{1pt}\setlength{\parsep}{0pt}

}

\end{document}